\documentclass[preprint,12pt]{elsarticle}

\usepackage{amssymb}
\usepackage{amsmath}
\usepackage{amsthm}
\usepackage{eucal}
\usepackage{mathrsfs}
\usepackage{subfigure}
\usepackage{fix-cm}
\usepackage{bm}
\usepackage{amsfonts}
\usepackage{enumerate}
\usepackage{enumitem} % enumerateのカスタマイズ
\usepackage{booktabs} % 
\usepackage{color}

\def \div{\mbox{\rm div}}
\def\x{x}

\journal{}

\begin{document}

\begin{frontmatter}



%\title{Topology optimization of viscoelastic microstructures with acoustic impedance manipulation}
\title{Topology optimization of isotropic viscoelastic microstructures based on periodic homogenization}


\author[inst1]{Hiroaki Deguchi}
\author[inst2]{Kei Matsushima}
\author[inst1,inst3]{Takayuki Yamada\corref{mycorrespondingauthor}}

\affiliation[inst1]{organization={Department of Mechanical Engineering, Graduate School of Engineering, The University of Tokyo},%Department and Organization
            addressline={Yayoi 2--11--16}, 
            city={Bunkyo--ku},
            postcode={113--8656}, 
            state={Tokyo},
            country={Japan}}
\affiliation[inst2]{organization={Informatics and Data Science Program, Graduate School of Advanced Science and Engineering, Hiroshima University},%Department and Organization
            addressline={Kagamiyama 1--7--1}, 
            city={Higashi--Hiroshima},
            postcode={739--8521}, 
            state={Hiroshima},
            country={Japan}}
\affiliation[inst3]{organization={Department of Strategic Studies, Institute of Engineering Innovation, Graduate School of Engineering, The University of Tokyo},%Department and Organization
            addressline={Yayoi 2--11--16}, 
            city={Bunkyo--ku},
            postcode={113--8656}, 
            state={Tokyo},
            country={Japan}}

\cortext[mycorrespondingauthor]{Corresponding author}
\ead{t.yamada@mech.t.u-tokyo.ac.jp}

\begin{abstract}
Mitigating low-frequency noise is particularly challenging due to its limited natural attenuation.
This study aims to design viscoelastic composite microstructures that achieve both low acoustic reflection and high internal damping by simultaneously enhancing their effective acoustic impedance and attenuation characteristics. 
Using complex-valued periodic homogenization theory and density-based topology optimization, viscoelastic and impedance-matching materials are designed within a highly symmetric unit cell to manipulate these isotropic properties. 
Numerical results show that the optimized isotropic design robustly outperforms its constituent materials and simple anisotropic laminate structures, exhibiting performance that is stable across a wide frequency band, independent of orientation, and stable across various structural scales.
This demonstrates the potential of microstructural engineering for effective low-frequency noise mitigation.

\end{abstract}




\end{frontmatter}


\section{Introduction}
\label{sec:intro}
Low-frequency noise, typically defined as sound with a frequency below 100 Hz, is a pervasive environmental issue originating from sources such as heavy machinery, transportation systems, and industrial activities \cite{leventhall2003review}. Its long wavelength allows it to propagate over extensive distances with low attenuation, leading to significant physical disturbances such as structural vibrations and adverse psychological and physiological effects on humans \cite{berglund2000new}. Traditional noise control strategies often rely on sound absorption, converting acoustic energy into heat, or sound insulation, which aims to block sound transmission \cite{alves2007vibroacoustic}.

This study specifically targets the challenge of mitigating low-frequency sound.
Impedance-matching materials, often realized as lightweight porous structures, play a crucial role by minimizing reflections at the material-air interface; achieving an acoustic impedance close to that of air is key to reducing this reflection \cite{kinsler2000fundamentals,cox2016acoustic,allard2009propagation}. However, for low-frequency sound, these materials typically require substantial thickness to be effective and may still suffer from significant sound transmission if they lack inherent dissipative mechanisms \cite{arenas2010recent}. While techniques such as adding air gaps or layering materials can enhance performance, they often introduce practical limitations, e.g., greater spatial requirements and complexity \cite{cox2016acoustic,arenas2010recent}. Another approach involves viscoelastic materials, which dissipate acoustic energy internally as heat due to their inherent damping properties, thus reducing transmitted sound \cite{lakes2009viscoelastic,ferry1980viscoelastic,nashif1991vibration}. Time-harmonic oscillation of viscoelastic materials is characterized by complex-valued, frequency-dependent elastic moduli, where the imaginary part (loss modulus) directly governs energy dissipation \cite{lakes2009viscoelastic,ferry1980viscoelastic,nashif1991vibration}. The challenge, however, is that viscoelastic materials often present a significant impedance mismatch with air, leading to strong reflections.

The inherent trade-off between the low reflection of impedance-matching materials and the high internal damping of viscoelastic materials motivates the exploration of viscoelastic composite microstructures. 
This research proposes that by strategically engineering the microscopic arrangement of these two constituent materials, it is possible to create a composite that synergistically integrates the low reflectivity of the impedance-matching phase and the high damping of the viscoelastic phase, leading to superior overall absorption performance while achieving robust performance at the macroscopic level.
To design such microstructures effectively, particularly for low frequencies where wavelengths are much larger than the microstructural scale, the periodic homogenization method based on two-scale asymptotic expansions provides a powerful analytical tool \cite{bensoussan2011asymptotic, sanchez-palencia1980non}. 
A notable advantage of this approach is its capability to determine effective material properties independent of the macroscopic scale without directly modeling the microscopic geometries.
Within this framework, employing a symmetric honeycomb lattice enables the design of macroscopically isotropic microstructures, thereby realizing several advantages over anisotropic designs: quantitative numerical characterization, orientation-independent performance, and stability across various structural scales.
This method allows for the derivation of effective macroscopic properties from the microstructural details, even when material properties are complex-valued, by formally extending the standard elastostatic framework \cite{allaire2012shape}.
Seminal work by \citet{yi1998asymptotic} specifically addressed the periodic homogenization of viscoelastic composites.

Topology optimization is a powerful tool for such systematic design \cite{bendsoe2003topology}.
This method optimizes material layout within a given design domain to achieve superior performance. Among various approaches, density-based methods, particularly the Solid Isotropic Material with Penalization (SIMP) method \cite{bendsoe1989optimal, bendsoe1999material}, have been widely adopted for their simplicity and effectiveness in diverse applications such as compliant mechanism design \cite{sigmund1997design,bruns2001topology,yamada2010topology,xia2016topology} and multiphysics systems \cite{sigmund2001design, tavakoli2014alternating,yoon2018multiphysics}. 
When combined with periodic homogenization, topology optimization becomes a promising strategy for designing material microstructures with desired macroscopic properties \cite{bendsoe1988generating, suzuki1991homogenization, sigmund1994materials,kim2020topology, coelho2015multiscale}. This multiscale approach has enabled the design of novel materials and metamaterials with extraordinary characteristics in acoustics \cite{noguchi2021topology, noguchi2021level}.
Non-gradient approaches based on the material-field series expansion (MFSE) method have also been applied to acoustic metamaterial design \cite{luo2020topology, zhang2023ultra, li2024topological, li2025customizable}.
Previous studies have explored topology optimization for maximizing damping in viscoelastic structures \cite{yun2017multi,zhang2021topology,matsushima2021topology} and have notably addressed the periodic homogenization of viscoelastic composites for optimization purposes \cite{huang2015topology,liu2018topology,giraldo2020fractional}, building upon a foundation of viscoelastic periodic homogenization \cite{yi1998asymptotic, gao2025dynamic}.
Extending these optimization frameworks to explicitly incorporate acoustic impedance control is therefore anticipated to enable the design of microstructures with enhanced acoustic performance, achieving both low reflection and high attenuation.

This study presents a topology optimization framework to design isotropic viscoelastic composite microstructures that minimize sound reflection while maximizing internal attenuation. 
An objective functional is formulated using the effective acoustic impedance and the imaginary part of the wavenumber, which are evaluated through periodic homogenization of the microstructure. 
The effectiveness of this approach is demonstrated through numerical examples. 
Subsequent validation confirms that the optimized isotropic design exhibits superior overall absorption compared to its constituent materials and is significantly more robust than simple anisotropic laminates, exhibiting performance that is both stable across a wide frequency band, independent of its orientation, and stable across various structural scales.
Ultimately, this work provides a systematic design framework for highly effective materials for low-frequency noise control.

The remainder of this paper is organized as follows. 
Section \ref{sec:model} describes the design problem, the governing equations for the acoustic and viscoelastic fields, and the derivation of effective properties using the periodic homogenization method. 
Section \ref{sec:topology optimization} explains the formulation of the topology optimization problem, the sensitivity analysis, and the optimization algorithm.
Section \ref{sec:numerical examples} presents the optimized microstructures and provides a detailed numerical validation of their superior and robust acoustic performance across various structural scales.
Finally, Section \ref{sec:conclusion} concludes the paper.

\section{Model}
\label{sec:model}

\subsection{Design problem for a low-reflection and high-damping structure}
\label{sec:design problem}
\begin{figure}%[ht]
	\begin{center}
	\includegraphics[scale=0.9]{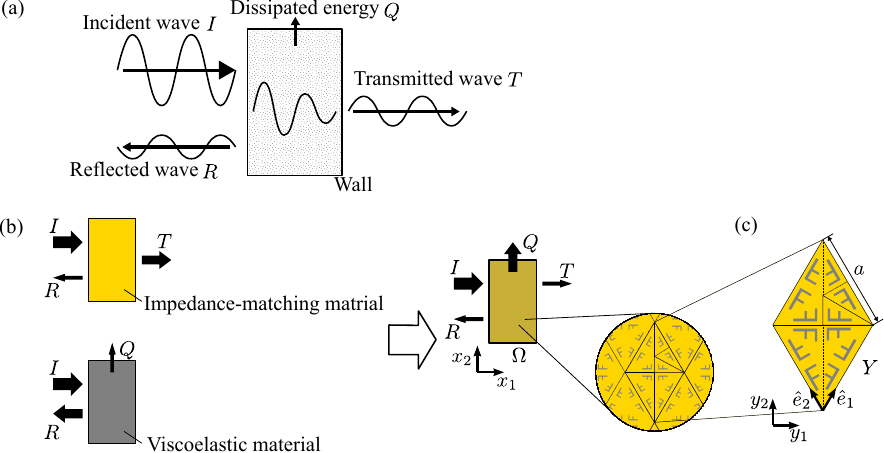}
	\caption{Model of a low-reflection and high-damping structure. (a) Illustration of the design problem. (b) Schematic of the viscoelastic composites. (c) The unit cell comprises two phases: impedance-matching material and viscoelastic material. The periodic structure is homogenized in the limiting case $a \to 0$.}
	\label{fig:model}
	\end{center}
\end{figure}

First, we consider the design of a structure that simultaneously exhibits low reflection and high damping characteristics for incident sound waves. Our aim is to minimize the energy reflected from the structure's front surface and the energy transmitted through its entirety.
As depicted in Fig.~\ref{fig:model}(a), an incident sound wave with energy $I$ impinges upon a structure. This interaction results in a reflected wave with energy $R$ and a transmitted wave with energy $T$. Within the structure, a portion of the sound energy, $Q$, is dissipated, typically as heat, due to the material's damping properties. Effective acoustic design, therefore, seeks to minimize the sum of the reflection coefficient ($R/I$) and the transmission coefficient ($T/I$).

In pursuit of this goal, a fundamental trade-off between low reflection and high damping often arises when considering single-phase materials. On one hand, an impedance-matching material can be designed to have an acoustic impedance close to that of the surrounding medium (e.g., air). As illustrated conceptually in Fig.~\ref{fig:model}(b) (top left), this minimizes the impedance mismatch at the interface, leading to a small reflection coefficient ($R/I$). However, such materials, if they are not inherently highly dissipative (e.g., lightweight porous materials), may allow a significant portion of the non-reflected energy to pass through, resulting in a large transmission coefficient ($T/I$).

On the other hand, a viscoelastic material is characterized by its ability to dissipate mechanical energy internally. As shown in Fig.~\ref{fig:model}(b) (bottom left), if a structure is made solely of a highly dissipative viscoelastic material, it can significantly attenuate the wave as it propagates, leading to a small transmission coefficient ($T/I$). However, viscoelastic materials often have acoustic impedances that are substantially different from that of air, causing a significant impedance mismatch at the incident surface and thus a large reflection coefficient ($R/I$).

This apparent conflict between minimizing reflection and maximizing internal attenuation suggests that neither material type alone is optimal for achieving both objectives simultaneously. Therefore, this study explores the potential of viscoelastic composite microstructures, as schematically shown in Fig.~\ref{fig:model}(b) (right). By strategically combining an impedance-matching material and a viscoelastic material at the microstructural level, we expect that it is possible to engineer a composite that synergistically leverages the strengths of both constituents. The aim is to create a material that not only presents a favorable impedance to the incident wave, thereby reducing reflections, but also possesses enhanced internal damping mechanisms to absorb the energy that does enter, thus minimizing transmission. The design of such a composite microstructure is the central focus of this work.

\subsection{Formulation of the design problem}
\label{sec:formulation}
We now derive the quantitative characterization of reflected and transmitted waves, which were discussed as the objectives in Section \ref{sec:design problem}.

\begin{figure}%[ht]
	\begin{center}
	\includegraphics[scale=1.0]{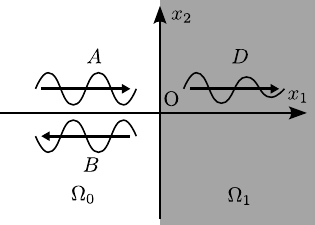}
	\caption{Half-plane filled with a viscoelastic material.}
	\label{fig:halfplane}
	\end{center}
\end{figure}

As shown in Fig.~\ref{fig:halfplane}, we consider the left half-plane $\Omega_0$ filled with air and the right half-plane $\Omega_1$ occupied by a viscoelastic material.

We assume that the acoustic pressure field $\tilde{p}(x,t)$ in the sound domain takes the time-harmonic form $\tilde{p}(x, t) = p(x)\exp(-\mathrm{i}\omega t)$, where $\omega$ is the angular frequency. Under this assumption, the governing equation of the acoustic pressure $p(x)$ in air is given by the following Helmholtz equation:
\begin{equation}
\nabla^2 p + k_0^2 p = 0 \quad \text{in } \Omega_0,
\label{eq:helmholtz_eq_sound}
\end{equation}
where $k_0 = \omega / c_0$ is the wavenumber in air, and $c_0 = \sqrt{\kappa_0/\rho_0}$ is the speed of sound, with $\kappa_0$ and $\rho_0$ denoting the bulk modulus and mass density of air, respectively.

Similarly, we assume that the displacement field $\tilde{u}(x,t)$ in the viscoelastic domain takes the time-harmonic form $\tilde{u}(x,t) = u(x)\exp(-\mathrm{i}\omega t)$. Under the plane-strain condition, the governing equation for the viscoelastic field can be expressed as
\begin{equation}
    \div(C : \varepsilon(u)) + \rho_1\omega^2u = 0 \quad \text{in } \Omega_1,
    \label{eq:helmholtz_eq_viscoelastic}
\end{equation}
where the elasticity tensor $C$ can be expressed using the Lam\'{e} constants $\lambda$ and $\mu$ as follows:
\begin{equation}
    C_{ijkl} = \lambda \delta_{ij}\delta_{kl} + \mu(\delta_{ik}\delta_{jl} + \delta_{il}\delta_{jk}),
    \label{eq:elasticity_tensor}
\end{equation}
where $\delta_{ij}$ is the Kronecker delta, and the Lam\'{e} constants are expressed using complex Young's modulus $E = E_1 + \mathrm{i}E_2$, with its imaginary part $E_2 \leq 0$ representing energy dissipation, and Poisson's ratio $\nu$ as follows:
\begin{equation}
    \lambda = \frac{E\nu}{(1+\nu)(1-2\nu)}, \quad \mu = \frac{E}{2(1+\nu)}.
    \label{eq:lame_constants}
\end{equation}
The strain tensor $\varepsilon(u)$ is related to the displacement $u$ as follows:
\begin{equation}
    \varepsilon_{ij}(u) = \frac{1}{2}\left( \frac{\partial u_j}{\partial x_i} + \frac{\partial u_i}{\partial x_j} \right).
    \label{eq:strain_tensor}
\end{equation}
The mass density of the viscoelastic material is denoted by $\rho_1$.
The stress tensor $\sigma$ is given by $\sigma_{ij} = C_{ijkl}\varepsilon_{kl}$.

Boundary conditions at the interface $\Gamma$ between air and the viscoelastic material are imposed as follows:
\begin{align}
    -p n &= \sigma \cdot n \quad \text{on } \Gamma \label{eq:bc1}, \\
    \frac{1}{\rho_0\omega^2}\nabla p \cdot n &= u \cdot n \quad \text{on } \Gamma \label{eq:bc2},
\end{align}
where $\bm{n}$ is the unit normal vector on the boundary $\Gamma$.
Without loss of generality, $\Gamma$ lies along the line $x_1=0$, and the unit normal vector $\bm{n}$ is chosen as $(1,0)$.

We consider an incident sound wave with amplitude $A$ propagating in the positive $x_1$ direction in $\Omega_0$, which generates a reflected sound wave with amplitude $B$ and a transmitted P-wave with amplitude $D$ in the viscoelastic material, as shown in Fig.~\ref{fig:halfplane}. Furthermore, we impose the radiation condition at a sufficiently far distance from the interface.
In this case, the sound pressure $p$ and the displacement field $\bm{u}$ that satisfy the governing equations (\ref{eq:helmholtz_eq_sound}) and (\ref{eq:helmholtz_eq_viscoelastic}) and the boundary conditions (\ref{eq:bc1}) and (\ref{eq:bc2}) can be expressed as follows:
\begin{align}
    p(x) &= A \exp \left( \mathrm{i}k_0x_1 \right) + B \exp \left( -\mathrm{i}k_0x_1 \right) \quad \text{in } \Omega_0, \\
    u_1(x) &= \mathrm{i} k_1 D \exp \left( \mathrm{i}k_1x_1 \right) \quad \text{in } \Omega_1, \label{eq:displacement1}\\
    u_2(x) &= 0 \quad \text{in } \Omega_1,
\end{align}
where $k_1 = \omega/c_1\in \mathbb{C}$ is the wavenumber, and $c_1 = \sqrt{(\lambda+2\mu)/\rho_1}$ is the speed of the P-wave in the viscoelastic material. 
From Eqs. \eqref{eq:bc1} and \eqref{eq:bc2}, the coefficients $A$, $B$, and $D$ satisfy the following relationships:
\begin{align}
    A + B &= \rho_1 \omega^2 D, \\
    A - B &= \rho_0 \omega^2 \frac{c_0}{c_1} D.
\end{align}
The reflection coefficient $R/I$ can be expressed as follows:
\begin{equation}
    \frac{R}{I} = \left|\frac{B}{A}\right|^2 = \left| \frac{\rho_1c_1 - \rho_0c_0}{\rho_1c_1 + \rho_0c_0} \right|^2. \label{eq:reflection_coefficient}
\end{equation}

The qualitative discussion in Section \ref{sec:design problem} regarding the objectives of low reflection and high attenuation can now be translated into quantitative targets.
First, we aim to minimize the reflection, which requires reducing the absolute difference between the acoustic impedances $\rho_1c_1$ and $\rho_0c_0$, in view of Eq. (\ref{eq:reflection_coefficient}).
Second, the attenuation characteristics of the viscoelastic material are enhanced when the imaginary part of the complex wavenumber $\mathrm{Im}[k_1]$ becomes larger, which is evident from Eq. (\ref{eq:displacement1}).

\subsection{Periodic homogenization for viscoelastic microstructures}
\label{sec:homogenization}
In this section, we employ a periodic homogenization method based on the two-scale asymptotic expansion technique to determine the effective properties of viscoelastic microstructures comprising a periodic lattice with two phases.
The viscoelastic nature of the materials implies that their mechanical properties, such as the elasticity tensor $C$ and Lam\'{e} parameters, are complex-valued and frequency-dependent. However, the mathematical framework of homogenization theory can be formally extended to such complex-valued coefficients, treating them similarly to real-valued coefficients in standard elastostatics \cite{allaire2012shape, akamatsu2024optimal}. This allows us to derive effective complex-valued homogenized properties for the composite.

The system of composite materials occupies a domain $\Omega$, which consists of a periodic array of unit cells.
In the region $\Omega$, the unit cell $Y$, a rhombus with side length $a$ defined by the translation vectors $a\hat{\mathbf{e}}_1$ and $a\hat{\mathbf{e}}_2$, where $\hat{\bm{e}}_1$ and $\hat{\bm{e}}_2$ are unit vectors in $\mathbb{R}^2$, is periodically arranged to form a honeycomb lattice, as shown in Fig.~\ref{fig:model}(b), (c).
Each unit cell is composed of two types of homogeneous materials: impedance-matching material and viscoelastic material.
Here, we consider a lattice structure with symmetry in the honeycomb lattice to ensure that the homogenized material tensor can be characterized as isotropic, described by only two independent complex Lam\'{e} constants \cite{tancogne2019stiffness}.
In fact, previous studies have demonstrated that optimal isotropic viscoelastic microstructures exhibit hexagonal symmetry even when designed on square domains with explicit constraints \cite{andreasen2014realization}.
This justifies the discussion in Section \ref{sec:formulation}. 
In addition, the macroscopic isotropy is highly desirable for ensuring performance independent of the material orientation.

In this system, the displacement field in the region $\Omega$ is represented by a solution of the elastodynamic problem, denoted by $u_a(x)$, as follows:
\begin{align}
    \left\{ 
    \begin{alignedat}{2}   
    \operatorname{div}\left(C\left(\frac{x}{a}\right) \nabla u_{a}\right) + \rho\left(\frac{x}{a}\right)\omega^2u_{a} &= f(x) \quad \mathrm{in} \quad \Omega, \\
    u_{a} &= 0 \quad \mathrm{on} \quad \partial{\Omega}, 
    \end{alignedat} 
    \right. \label{eq:dynamic_elasticity}
\end{align}
where the variable $x \in \mathbb{R}^2$ represents the macro-scale coordinates, $y=x/a$ represents the micro-scale coordinates normalized by the unit cell size $a$, and $f(x)$ represents the source term.
The fourth-order elasticity tensor $C(y)$ and mass density $\rho(y)$ are $Y$-periodic. Namely, for the basis vectors $\hat{e}_1$ and $\hat{e}_2$, the coefficients satisfy the following periodicity condition:
\begin{align}
    C(y+\hat{e}_i) &= C(y), \quad \rho(y+\hat{e}_i) = \rho(y) \quad (i=1,2). \label{eq:periodicity}
\end{align}

We use a periodic homogenization method to characterize the periodic array as a homogenized viscoelastic microstructure. 
The principle of this method is that for a sufficiently small $a$, the solution $u_a$ to Eq. \eqref{eq:dynamic_elasticity} can be approximated by the solution $u$ of the following homogenized problem:
\begin{align}
    \left\{ 
    \begin{alignedat}{2}   
    \operatorname{div}\left(C^H \nabla u\right) + \rho^H\omega^2u &= f \quad \mathrm{in} \quad \Omega, \\
    u &= 0 \quad \mathrm{on} \quad \partial{\Omega}, 
    \end{alignedat} 
    \right. \label{eq:homogenizaed_problem}
\end{align}
where the constant tensors $C^H$ and $\rho^H$ are known as the homogenized elasticity tensor and mass density, respectively \cite{allaire2012shape}. When the unit cell size $a$ is sufficiently small compared to the wavelength, the homogenized elasticity tensor $C^H_{ijkl}$ and mass density $\rho^H$ can be characterized as follows \cite{allaire2012shape, akamatsu2024optimal}:
\begin{align}
    C^H_{ijkl} &= \frac{1}{|Y|}\int_Y \left(g_{ij} + \varepsilon(\chi_{ij})\right):C:g_{kl} \mathrm{d}y, \\
    \rho^H &= \frac{1}{|Y|}\int_Y \rho(y) \mathrm{d}y,
\end{align}
where $|Y|$ is the volume of the unit cell, and the second-order tensor $g$ is defined by $g_{ij} = \frac{1}{2}\left(\hat{e}_i\otimes\hat{e}_j + \hat{e}_j\otimes\hat{e}_i\right)$, and $\chi_{ij}$ are solutions of the following unit cell problem:
\begin{align}
    \left\{ 
    \begin{alignedat}{2}   
    -\operatorname{div}\left(C: \varepsilon(\chi_{ij})\right) = \div\left(C: g_{ij}\right) \quad &\mathrm{in} \quad Y, \\
    y \mapsto \chi_{ij} (y) \quad &Y\text{-periodic}
    \end{alignedat}
    \right. \label{eq:unit_cell_problem}
\end{align}
for each $i,j=1,2$.

\section{Topology optimization}
\label{sec:topology optimization}
This section is devoted to a density-based topology optimization method for the design problem described in Section \ref{sec:design problem}.
We first formulate our topology optimization problem and subsequently explain its sensitivity analysis and the optimization algorithm.

\subsection{Formulation of topology optimization problem}
\label{sec:topologyoptimization problem}
A two-phase topology optimization based on the Solid Isotropic Material with Penalization (SIMP) method \cite{bendsoe1999material, tavakoli2014alternating} is used to solve a structural optimization problem in the unit cell $Y$. A density function $\theta:Y\to [0,1]$ is defined in the unit cell $Y$ to represent the material distribution of the viscoelastic material. The elasticity tensor $C(y)$ and mass density $\rho(y)$, which depend on the density function $\theta$, are expressed as follows:
\begin{align}
    C(y) &= \theta(y)^3 C_1 + (1-\theta(y)^3)C_2, \\
    \rho(y) &= \theta(y)^3 \rho_1 + (1-\theta(y)^3)\rho_2,
\end{align}
where $C_1$ and $\rho_1$ are the elasticity tensor and mass density of the viscoelastic material, and $C_2$ and $\rho_2$ are those of the impedance-matching material.

The objective is to find a density function $\theta$ such that it minimizes the reflection coefficient $R/I$ and maximizes the imaginary part of the complex wavenumber $\mathrm{Im}[k_1]$. The objective functional $J(\theta, \chi(\theta))$ is defined as follows:
\begin{align}
    J(\theta, \chi(\theta)) &= b_1 J_1(\theta, \chi(\theta)) + b_2 J_2(\theta, \chi(\theta)), \\
    J_1(\theta, \chi(\theta)) &= \left(1 - \left|1-\frac{\sqrt{\kappa_0\rho_0}}{\sqrt{(\lambda^{H}_{\chi(\theta)}+2\mu^{H}_{\chi(\theta)})\rho^{H}_{\theta}} }\right|^2 \right)^2, \\
    J_2(\theta, \chi(\theta)) &= \sqrt{\frac{\kappa_0}{\rho_0}} \mathrm{Im} \left[ \sqrt{\frac{\rho^{H}_{\theta}}{\lambda^{H}_{\chi(\theta)}+2\mu^{H}_{\chi(\theta)}}} \right],
\end{align}
where $b_1$ and $b_2$ are the weights of the objective functionals, $\lambda^{H}$ and $\mu^{H}$ are the Lam\'{e} constants derived from the homogenized elasticity tensor $C^H$, and $\rho^{H}$ is the homogenized density. 
The first term $J_1$ is formulated to promote impedance matching, and the second term $J_2$ represents a scaled version of the imaginary part of the complex wavenumber in the homogenized viscoelastic medium, which is associated with attenuation.

The optimization problem can be expressed as follows:
\begin{align}
    \max_{\theta} \quad & J(\theta, \chi(\theta)).
\end{align}

\subsection{Sensitivity analysis}
\label{sec:sensitivity analysis}
The derivative of the objective functional $J(\theta, \chi(\theta))$ with respect to the density function $\theta$ is called the sensitivity of the objective functional. It is calculated using the adjoint variable method \cite{akamatsu2024optimal}. As discussed in Section \ref{sec:homogenization}, the objective functional $\theta \mapsto J(\theta, \chi(\theta))$ is expressed as a function of the homogenized elasticity tensor $C^H_{\chi(\theta)}$ and mass density $\rho^H_{\theta}$; thus, it is necessary to calculate the derivatives of these homogenized properties with respect to the density function $\theta$. Using the adjoint variable method, the derivatives of the homogenized elasticity tensor and mass density with respect to $\theta$ can be obtained as follows \cite{akamatsu2024optimal}:
\begin{align}
    \mathrm{d} C^{H}_{i j k l}(\chi(\theta))[\tilde{\theta}] &= \frac{1}{|Y|} \int_Y\left(g_{i j}+\varepsilon\left(\chi_{i j}(\theta) \right)\right): \mathrm{d} C(\theta)[\tilde{\theta}]: (g_{k l}+\varepsilon\left(\chi_{k l}(\theta)\right)) \mathrm{d} y, \label{eq:Derivative of C} \\ 
    \mathrm{d} \rho^{H}_{\theta}[\tilde{\theta}]&= \frac{1}{|Y|} \int_Y \mathrm{d} \rho_\theta[\tilde{\theta}] \mathrm{d}y, \label{eq:Derivative of rho}
\end{align}
where $\mathrm{d} F(\theta)[\tilde{\theta}]$ denotes the derivative of a functional $F$ with respect to the direction $\tilde{\theta}$ at $\theta$, i.e., 
\begin{align}
    F(\theta+\tilde{\theta}) = F(\theta) + \mathrm{d} F(\theta)[\tilde{\theta}] + o(\|\tilde{\theta}\|).
\end{align}

\subsection{Optimization algorithm}
\label{sec:optimization algorithm}
\begin{figure}%[ht]
	\begin{center}
	\includegraphics[scale=1.0]{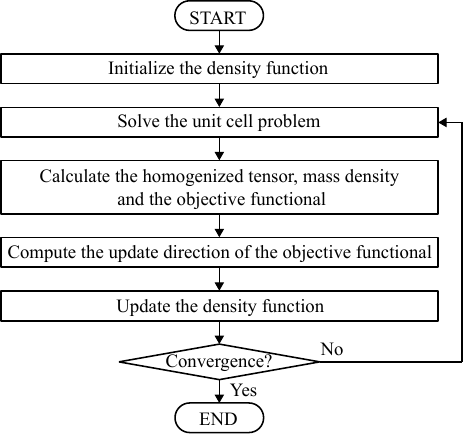}
	\caption{Flowchart of the optimization method.}
	\label{fig:algorighm}
	\end{center}
\end{figure}

The optimization algorithm is shown in Fig.~\ref{fig:algorighm}.
The detailed steps are as follows:
\begin{enumerate}[label=step \arabic* :, leftmargin=*, align=left]
    \item Initialize the density function $\theta$.
    \item Solve the unit cell problem (\ref{eq:unit_cell_problem}) to obtain $\chi_{ij}(\theta)$.
    \item Calculate the homogenized elasticity tensor $C^H$, mass density $\rho^H$, and the objective functional $J(\theta, \chi(\theta))$.
    \item Compute the update direction $\tilde{\theta}$ of the objective functional $J(\theta, \chi(\theta))$ using the derivatives obtained from Eqs. \eqref{eq:Derivative of C} and \eqref{eq:Derivative of rho} via the chain rule.
    \item Update the density function as $\theta \leftarrow \theta + \alpha \tilde{\theta}$, where $\alpha$ is the step size.
    \item Return to step 2 until the convergence condition is satisfied.
\end{enumerate}

The open-source finite element method solver FreeFEM++ \cite{hecht2012new} is used to perform the finite element analysis and sensitivity calculations.

\section{Numerical examples}
\label{sec:numerical examples}
In this section, we present numerical examples to demonstrate the effectiveness of the proposed method. For the constituent materials, we chose low-density polyethylene (LDPE) as the viscoelastic material for its large loss tangent \cite{hay2013measuring} and lightweight porous melamine foam as the impedance-matching material \cite{bonfiglio2015determination}. The physical properties used in the simulations are listed in Table \ref{tab:parameters}. The complex-valued Young's moduli of both materials are frequency-dependent; the values listed in the table were adopted from experimental results at a frequency of approximately $f:=\omega/(2\pi) = 100$ Hz \cite{hay2013measuring,bonfiglio2015determination}. The mass density of air is set to $\rho_0 = 1.2$ $\mathrm{kg/m^3}$, and its bulk modulus is $\kappa_0 = 1.4 \times 10^5$ $\mathrm{Pa}$.

\begin{table}[htbp]
    \centering
    \caption{Physical properties of each material.}
    \begin{tabular}{lcccc}
    \toprule
    Material & $f$ (Hz) & E (Pa) & $\nu$ & $\rho$ (kg/m$^3$) \\
    \midrule
    LDPE      & 100 & $1.1 \times 10^{9} - 2.0 \times 10^{8}\mathrm{i}$ & 0.33 & $1.2 \times 10^{3}$ \\
    melamine foam & 100 & $9.0 \times 10^{4} - 1.1 \times 10^{3}\mathrm{i}$ & 0.12 & 9.0 \\
    \bottomrule
    \end{tabular}
    \label{tab:parameters}
\end{table}

\subsection{Optimized design}
\label{sec:optimized design}
\begin{figure}[htbp]
	\begin{center}
	\includegraphics[width=0.8\linewidth]{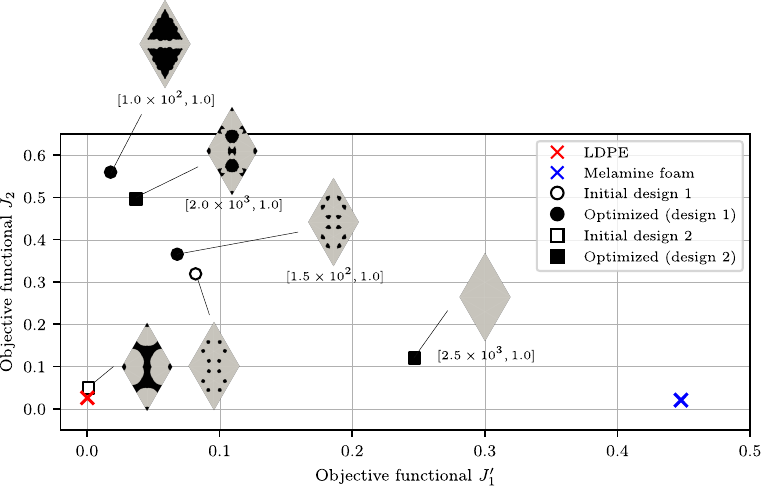}
	\caption{Distribution of optimized designs in the objective function space ($J'_1$--$J_2$). The markers indicate the results for different weight ratios and initial designs, compared with the constituent materials (LDPE and melamine foam).}
    \label{fig:pareto}
	\end{center}
\end{figure}

In this study, optimizations were performed under several conditions to investigate the influence of the weight coefficients $b_1, b_2$, and the initial density distribution on the optimization results. 
Fig.~\ref{fig:pareto} shows the distribution of the optimized designs in the objective function space ($J'_1$--$J_2$). 
Since the values of the impedance-matching functional $J_1$ tend to be extremely small, we plotted the results using $J'_1 := \sqrt{J_1}$ to enhance the visual clarity of the distribution. 
Since the objective functionals $J_1$ and $J_2$ are formulated to be dimensionless, the weights $b_1$ and $b_2$ are also dimensionless.
Regarding the weight coefficients, we first experimentally determined the reference values $\bar{b}_1=5.0\times 10^{-4}$ and $\bar{b}_2 = 10.0$ that ensure the optimization progresses appropriately when one of the coefficients is set to zero. Then, we varied the weights by multiplying these references by the scaling factors $[\tilde{b}_1, \tilde{b}_2]$. 
This systematic variation of the weight ratio allowed for a comprehensive investigation of the correlation between the objective functionals. 
To examine the robustness of the design exploration, two different initial configurations were employed: Initial design 1, with a viscoelastic body of radius $3a/80$ at the unit cell, and Initial design 2, with a melamine foam of radius $3a/16$. 
Additionally, the lower bound of the density function $\theta$ is set to 0.05 to ensure numerical stability throughout the process.
As illustrated in Fig.~\ref{fig:pareto}, although the optimization converges to different solutions depending on the initial density distribution, both cases yield high-performance designs.

The distribution in Fig.~\ref{fig:pareto} reveals a trade-off between the impedance matching $J_1$ and the attenuation performance $J_2$. 
While the performance of the composite material regarding $J_1$ does not exceed that of the pure melamine foam, the optimized designs significantly outperform the pure LDPE in terms of the attenuation functional $J_2$.
Notably, for Initial design 2 with large $\tilde{b}_1$, $\theta$ reaches its lower bound of 0.05. This result differs from pure melamine foam because the non-zero lower bound still contributes to the homogenized properties.

\begin{figure}%[ht]
	\begin{center}
	\includegraphics[scale=0.8]{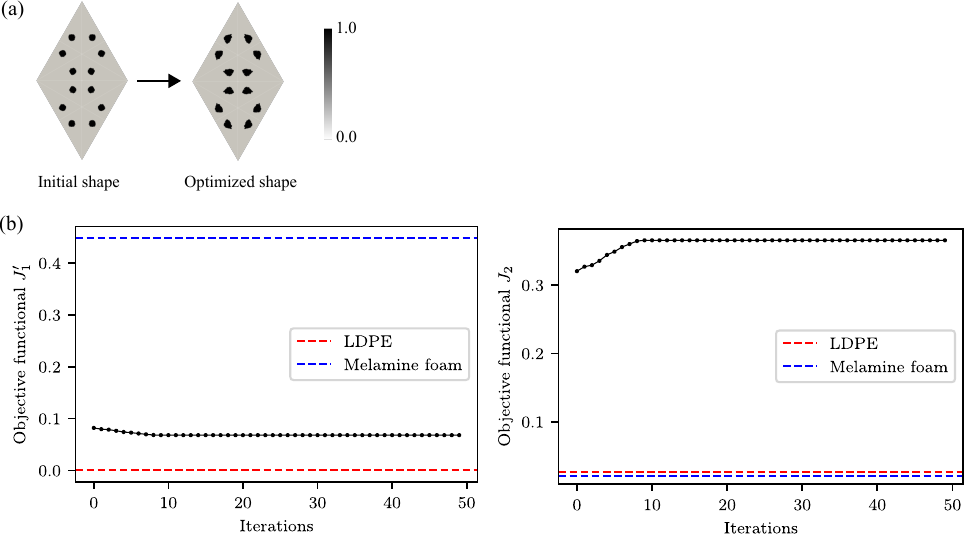}
	\caption{Histories of the optimized design process. (a) Initial and optimized shapes.
    (b) History of the objective functional $J'_1:= \sqrt{J_1}$ and $J_2$. 
    }	
    \label{fig:history}
	\end{center}
\end{figure}

In the following discussion, we focus on the optimized design obtained with the weight coefficients $[\tilde{b}_1, \tilde{b}_2] = [1.5\times 10^2, 1.0]$ and Initial design 1, which represents a well-balanced performance between the two objectives. 
Fig.~\ref{fig:history} shows the results of the topology optimization for this representative case.

To verify the macroscopic isotropy of the obtained microstructure, the homogenized elasticity tensor $C^H$ calculated from the optimized unit cell is presented as follows:
\begin{align}
    &C^H = 
    \begin{bmatrix}
    C_{1111}^H & C_{1122}^H & C_{1112}^H \\
    C_{2211}^H & C_{2222}^H & C_{2212}^H \\
    C_{1211}^H & C_{1222}^H & C_{1212}^H
    \end{bmatrix} \\
    &= \Biggl[
    \begin{array}{cc}
    3.51 \times 10^5 - 4.44 \times 10^4\mathrm{i} & 1.31 \times 10^5 - 2.08 \times 10^4\mathrm{i} \\
    1.31 \times 10^5 - 2.08 \times 10^4\mathrm{i} & 3.51 \times 10^5 - 4.44 \times 10^4\mathrm{i} \\
    -2.63 \times 10^{-11} + 1.86 \times 10^{-10}\mathrm{i} & 8.47 \times 10^{-11} - 1.45 \times 10^{-10}\mathrm{i}
    \end{array}
    \Biggr. \nonumber \\
    &\hspace{5em}
    \Biggl.
    \begin{array}{c}
    -2.63 \times 10^{-11} + 1.86 \times 10^{-10}\mathrm{i} \\
    8.47 \times 10^{-11} - 1.45 \times 10^{-10}\mathrm{i} \\
    1.10 \times 10^5 - 1.18 \times 10^4\mathrm{i}
    \end{array}
    \Biggr]
    \label{eq:homogenized_tensor}
\end{align}
The calculated results satisfy the conditions $C_{1111}^H \approx C_{2222}^H$ and $2C_{1212}^H \approx C_{1111}^H - C_{1122}^H$, which confirms that the homogenized elasticity tensor $C^H$ fulfills the requirement for macroscopic isotropy.

\subsection{Validation of the optimized design}
\label{sec:validation}
In this section, the acoustic performance of the optimized microstructure obtained by the proposed method is validated using numerical simulations. For this purpose, we employ the Transfer Matrix Method (TMM) and the Finite Element Method (FEM). For comparison, the performance of homogeneous materials composed of the constituent materials (LDPE and melamine foam) and simple anisotropic laminate structures is also evaluated.

\begin{figure}%[ht]
	\begin{center}
	\includegraphics[scale=0.52]{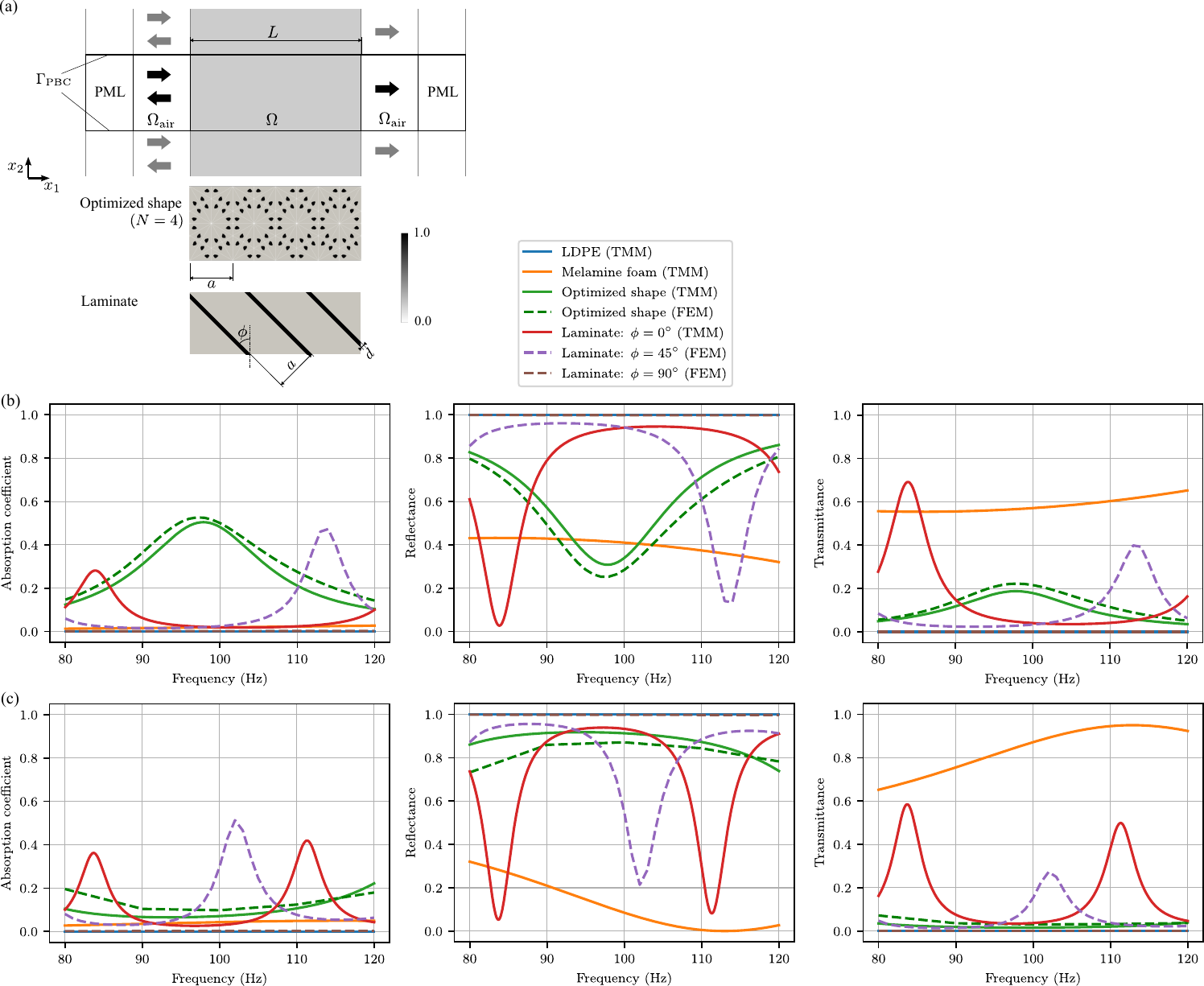}
	\caption{
        Schematic of the verification method and results.
        (a) $x_2$-periodic computational model and boundary conditions for verification. $N$ layers of unit cells are used to compare the performance of the optimized design and the $\phi$-tilted laminate structure.
        (b) and (c) Comparison of absorption, transmission, and reflection coefficients for wall thicknesses $L=0.3$ m (approx. $1/2$ wavelength) and $L=0.45$ m (approx. $3/4$ wavelength), respectively.
        }	
    \label{fig:verification}
	\end{center}
\end{figure}

The schematic of the validation model is shown in Fig.~\ref{fig:verification}(a). 
To investigate the dependence of acoustic characteristics on the structural scale, we consider two wall thicknesses: $L=0.3$ m and $L=0.45$ m. These values correspond approximately to 1/2 and 3/4 of the longitudinal wavelength at 100 Hz, representing the resonance thickness for peak absorption and the thickness at which performance is relatively lowest, respectively.

A plane wave with a frequency band of 90 Hz to 110 Hz is incident on this wall from the $x_1$ direction, and the reflected wave energy $R$ and transmitted wave energy $T$ are calculated. The performance is evaluated using the absorption coefficient $1-R/I-T/I$, the transmission coefficient $T/I$, and the reflection coefficient $R/I$, where $I$ is the incident wave energy.

The acoustic performance of the homogeneous materials and the laminate structures layered in the $x_1$ direction was calculated using TMM. On the other hand, the performance of the wall with the optimized microstructure was analyzed using the FEM model shown in Fig.~\ref{fig:verification}(a). In this model, periodic boundary conditions were applied to the $\Gamma_\mathrm{PBC}$ boundaries along the $x_2$ direction. Additionally, Perfectly Matched Layers (PML) were placed on the incident and transmission sides to suppress reflections. The optimized structure was modeled by stacking $N=32$ unit cells in the $x_1$ direction. For the comparative laminate structures, the thickness $d$ of the LDPE layer was determined to match the volume fraction of the optimized structure, and their performance was evaluated for three laminate angles: $\phi=0^\circ, 45^\circ, 90^\circ$.

Figs.~\ref{fig:verification}(b) and (c) show the numerical results for the thicknesses $L=0.3$ m and $L=0.45$ m, respectively. Three key findings emerge from these results.
First, the validity of the homogenization approach used in this study is confirmed. The performance calculated by TMM using homogenized properties shows excellent agreement with the results from the direct FEM analysis of the $N=32$ layer unit cell structure.
Second, the optimized structure exhibits significantly superior performance compared to its constituent materials alone across both thickness conditions. 
While the pure LDPE wall exhibits high reflection due to impedance mismatch and the pure melamine foam allows high transmission, the optimized microstructure effectively suppresses both reflection and transmission by synergizing impedance matching with internal dissipation.
The absolute level of absorption is, however, inherently bounded by the trade-off between impedance matching and internal damping in microstructural optimization alone, as discussed in Section~\ref{sec:design problem} and evidenced by the Pareto distribution in Fig.~\ref{fig:pareto}. Nevertheless, the composite microstructure achieves absorption performance far superior to that of either constituent material.
Third, the optimized structure demonstrates more robust sound absorption than simple anisotropic laminate structures, particularly with respect to the frequency of the incident wave and the orientation of the microstructure.
Laminate structures exhibit good performance only within a limited range of frequencies and angles (e.g., $\phi=0^\circ$). 
The optimized microstructure, by contrast, maintains superior performance consistently across the entire evaluated frequency band; furthermore, because it behaves as an isotropic material on a macroscopic scale, its performance is independent of the microstructure's orientation.

These validations demonstrate that the proposed optimization method is effective for designing high-performance acoustic materials that satisfy the conflicting requirements of low reflection and high attenuation.
Specifically, by optimizing the microscopic composite structure, we achieved macroscopically robust performance that remains stable across various structural scales and orientations.

\section{Conclusion}
\label{sec:conclusion}
In this study, we proposed a topology optimization method for isotropic viscoelastic composites to minimize both reflected and transmitted waves. We formulated an objective functional based on effective acoustic impedance and attenuation, which were efficiently evaluated using a complex-valued periodic homogenization method. This approach was integrated into a density-based optimization framework, with design sensitivities calculated via the adjoint variable method.

Numerical examples demonstrated the effectiveness of the method. The optimized structure achieved attenuation characteristics superior to the pure viscoelastic material while maintaining low reflection. 
Subsequent numerical validation confirmed the accuracy of the periodic homogenization approach.
The validation also showed that the optimized design robustly outperforms its constituent materials and simple anisotropic laminate structures, as it maintains high performance across the evaluated frequency band and exhibits stable performance across various structural scales and orientations, owing to the microstructural optimization and the resulting macroscopic isotropy.

We conclude that the proposed method is effective for designing high-performance isotropic viscoelastic composites.
The results demonstrate that the optimization of the microscopic composite structure is key to achieving macroscopically robust acoustic performance.
Beyond this microstructural scale, concurrent macro-scale design and appropriate material selection represent promising directions for future research to further improve the absorption coefficient.

\section*{CRediT authorship contribution statement}
\textbf{Hiroaki Deguchi:} Writing -- original draft, Visualization, Validation, Formal analysis, Data curation. 
\textbf{Kei Matsushima:} Writing -- review \& editing. 
\textbf{Takayuki Yamada:} Supervision.

\section*{Declaration of Generative AI and AI-assisted technologies in the writing process}
During the preparation of this work, the authors used Gemini in order to refine the language and improve the clarity of the manuscript. After using this tool, the authors reviewed and edited the content as needed and take full responsibility for the content of the published article.

\section*{Declaration of competing interest}
The authors declare that they have no known competing financial interests or personal relationships that could have appeared to influence the work reported in this paper.

\section*{Acknowledgements}
This work was supported in part by JSPS KAKENHI Grant Number JP23H03800 and JST FOREST Program Grant Number JPMJFR202J.

\section*{Data availability}
Data will be made available on request.

%%       \citet{<label>} ==> Jones et al. [21]
%%       \citep{<label>} ==> [21]
%%

%Level set-based topology optimization presented by \citet{yamada2010topology} is used.


\bibliographystyle{elsarticle-num-names} 
\bibliography{mybibfile}





% \end{thebibliography}
\end{document}